\documentclass[12pt,a4paper]{article}
\usepackage{graphicx}
\usepackage{epsfig}

\begin{document}

\title {How much multifractality is included in monofractal signals?}

\author {Dariusz Grech\footnote{dgrech@ift.uni.wroc.pl} and Grzegorz Pamu{\l}a\footnote{gpamula@ift.uni.wroc.pl} \\
Institute of Theoretical Physics\\
University of Wroc{\l}aw, PL-50-204 Wroc{\l}aw, Poland}
\date{}

\maketitle

\begin{abstract}

We investigate the presence of residual multifractal background  for monofractal signals which appears due to the finite length of the signals and (or) due to the long memory the signals reveal. This phenomenon is investigated numerically within the multifractal detrended fluctuation analysis (MF-DFA) for artificially generated data. Next, the analytical formulas enabling to describe the multifractal content in such signals are provided. Final results are shown in the frequently used generalized Hurst exponent $h(q)$ multifractal scenario and are presented as a function of time series length $L$ and the autocorrelation exponent value $\gamma$. The multifractal spectrum $(\alpha, f(\alpha))$ approach is also discussed. The obtained results may be significant in any practical application of multifractality, including financial data analysis, because the 'true' multifractal effect should be clearly separated from the so called 'multifractal noise'.
Examples from finance in this context are given. The provided formulas may help to decide whether we do deal in particular case with the signal of real multifractal origin. They also push  further findings already existing in literature.
\end{abstract}

$$
$$

\vspace{1cm}

\section{Introduction}

Multifractality \cite{mf12,mf13,mf14,kantelhardt-preprint,mf1,mf2,mf3} is the property of complex and composite systems that has been attracting more and more attention in recent years. The practical fruits of multifractality are not precisely  known  yet but at least in the case of financial markets some interesting features of this phenomenon were shown (see e.g. \cite{mf4,mf5,mf6,mf7,mf8,mf9,mf10,mf11,oswiecimka,czarnecki}) that rise hopes for future applications. Since the paper by Kantelhardt {\em et.al.} \cite{Kantelhardt-316} we know that multifractality may result not only from long-range correlations but also from fat tails in probability distributions (PDF) of investigated data. Normally, one expects multifractality in time series as a result of different form of autocorrelations appearing at various time scales. However, we always deal in practise with finite samples of data collected in time series of given length. In such a case, multifractality may appear even if no difference in autocorrelation properties exists for various time scales. It is because large fluctuations cannot be detected as frequent as small fluctuations in finite samples of data with long memory -- mainly due to the insufficient data statistics. In other words, large fluctuations are not able to be formed in small samples of data, contrary to small fluctuations. Therefore, one gets in the case of shorter time series the multifractal property which itself is not programmed to be multifractal in a sense of different autocorrelation properties at various scales. The latter multifractality,  related to variety of autocorrelations, is more substantial and has to be somehow separated from the former one which we shall call the multifractal background or residual noise further on. The preliminary analysis of this problem had already been made in \cite{zhou,drozdz}. Our goal is to describe the presence of multifractal background quantitatively for time series with long memory induced by explicit form of autocorrelations in data. Our approach is directly based on Fourier filtering method (FFM) \cite{ffm} and differs therefore from other approaches, where the long-memory effect was inserted into a signal by the particular choice of power spectrum (amplitude adjusted Fourier transform method \cite{IAAFT}) \cite{zhou} or log-normal cascade implementation \cite{drozdz}, instead of direct shaping the artificial data with autocorrelation exponent $\gamma$ discussed further on.

We use the multifractal detrended fluctuation analysis (MF-DFA) \cite{kantelhardt-preprint} as the commonly accepted technique to find multifractal properties of  time series. This method is described elsewhere (see e.g. \cite{kantelhardt-preprint,oswiecimka,czarnecki,Kantelhardt-316}) so we will not recall it in details here. To keep the standard notation we will note the $q$-deformed fluctuation of the time series signal around its local trend  (assumed linear in our approach) in a time window of size $\tau$ as $F_q(\tau)$. Usually, the multifractal properties are presented as the multifractal spectrum $(\alpha, f(\alpha))$ \cite{mf3}, called sometimes also H{\"o}lder description. Equivalently, in the Hurst language, one can consider the spread of generalized Hurst exponents $h(q)$ \cite{Kantelhardt-316}, calculated within MF-DFA for $q-\mathrm{deformed}$ fluctuations $F_q(\tau)$ from the power law:
\begin{equation}
F_q(\tau) \sim \tau^{h(q)}
\end{equation}
Both descriptions are linked together via relations \cite{legendre1,legendre2}
\begin{equation}
\alpha=h(q)+qh'(q),\qquad f(\alpha) = q(\alpha-h(q))+1
\end{equation}

We start in this article with the generalized Hurst exponent $h(q)$ description of multifractality. Our results are then easy translated into  H{\"o}lder language with the use of Eq.(2). Finally, we compare our findings with the properties of real data from financial market, commonly believed to exhibit multifractal features.

\section{Generalized Hurst exponents for finite monofractal signals}

Our aim is to evaluate the multifractal effect in finite artificial signals of various lengths for given constant level of persistency, i.e. that autocorrelations are not changing with the time scale. Such signals are built by us within FFM \cite{ffm}. The level of autocorrelations is modulated by the proper choice of scaling exponent $\gamma$ responsible for the magnitude of autocorrelation function $C(\tau)$. The latter one satisfies for stationary series with long memory the known power law:
\begin{equation}
C(\tau)=\langle \Delta x(t)\Delta x(t+\tau) \rangle \sim \tau^{-\gamma}
\label{corr}
\end{equation}
where $\Delta x(t)=x(t+1)-x(t)$ are increments of time series, $\tau$ is the time-lag between observations and the average $\langle\rangle$  is taken over all data in the series.

The $\gamma$ scaling exponent may be linked to the Hurst exponent $H$ \cite{hurst} by the formula \cite{Kantelhardt-295}:
\begin{equation}
 \gamma=2-2H.
\end{equation}

In the quantitative analysis of residual multifractality left in monofractal finite signals, we concerned the ensembles of numerically generated time series of length $L=2^n$, $(n=9, 10,\ldots, 20)$ with the pre-assumed autocorrelation exponent value $\gamma = 0.1, 0.2,\ldots, 0.9, 1.0$, each containing $10^2$ independent realizations. Thus, the spread of $\gamma$ exponents covers the range $1/2\leq H<1$. The obtained quantities have been averaged over such statistical ensemble with given $L$ and $\gamma$ as input parameters.

First, we examined the FFM procedure for time series generation, in order to check its accuracy towards replication of the pre-assumed autocorrelation properties coming out from the particular choice of $\gamma$ exponent as input. Fig.1 demonstrates its efficiency. It is seen that the power law in Eq.(3) is reproduced very well even for large time-lags. Moreover,
a coincidence between input and output $\gamma$'s is also satisfactory. The length of generated data-samples was chosen as  powers of $2$ to improve performance of fast Fourier transform algorithm.

The next problem we had to examine, is the performance of MF-DFA technique which strictly depends on the power law scaling  between  $q-\mathrm{deformed}$ fluctuations $F_q(\tau)$ and the box size $\tau$ (see Eq.(1)). An exact extraction of the generalized Hurst exponent $h(q)$ is then possible only for well determined scaling range in the fitting procedure $\log F_q(\tau)$ vs $\log \tau$. Fig.2 clearly shows the expected power law dependence for various lengths of the signal $L$ and for different values of deformation parameter $q$. The latter one was uniformly distributed in the range from $-15$ to $+15$. These plots justify the scaling range from $\tau=10$ till $ \tau=L/4$ which was chosen by us to be used further on.

To determine quantitatively  the amount of multifractal residual noise present in given time series, the edge values of $h(q)$ function were investigated for them. Let us introduce the new parameter $\Delta h$ defined as the difference between the two asymptotic limits
\begin{equation}
\Delta h = \lim_{q\to-\infty} h(q) - \lim_{q\to\infty} h(q)
\end{equation}
and assume for numerical reasons that such asymptotic limits are reached already at $q=\pm 15$. Such assumption is justified in Fig.3, where plots for $h(q)$ are shown for $L=2^{12}, 2^{20}$ and for $\gamma = 0.1, 0.5, 0.9$ respectively.

Generally, we may expect that $\Delta h$ is the unknown function of $L$ and $\gamma$. The form of $\Delta h(L, \gamma)$ dependence is thus a crucial problem. To simplify it, one may consider the case $\gamma = \mathrm{const}$ for a moment, e.g. $\gamma=1$ ($H=1/2$), corresponding to uncorrelated data. Fig.4 shows the edge characteristics of $h(q)$ presented for two distinct series of length $L=2^{12}, 2^{20}$, generated with the autocorrelation exponent $\gamma$, and then shuffled. The dependence on $\gamma$ is evidently absent proving that shuffling procedure was effective enough, while the residual dependence  $\Delta h(\pm 15)$ on $L$ is still kept and obvious.

The detailed analysis of the latter dependence is revealed in Fig.5 collecting results for various data lengths. Astonishingly, this figure suggests a power law dependence between \linebreak\mbox{$\Delta h_1 \equiv \Delta h(\gamma=1)$} and $L$.

\begin{equation}
\Delta h_1(L) = C_1L^{-\eta_1},
\label{bc1}
\end{equation}
where $C_{1}$ and $\eta_1$ are constant.

The knowledge of $95\%$ confidence level for this relation is crucial in practise. Its meaning is that any result measured above the particular value has probability less than $5\%$. To obtain this confidence level one has to correct $C_1$ and $\eta_1$ parameters by the corresponding quantiles calculated from the $1\sigma$ uncertainties $\sigma_C$,  $\sigma_{\eta}$ of the fit and from the standard deviation $S$ resulting from the series statistics\footnote{exponential dependence in this formula comes from the uncertainty of regression fit in logarithmic scale}:
 \begin{equation}
\Delta h_1^{95\%}(L) = C_1\exp(f(\sigma_C+S))L^{-\eta_1+f\sigma_\eta}.
\label{bc1}
\end{equation}
where $f=1.65$ is the respective factor for the particular $95\%$ confidence level.

Let us take now a closer look at the case of autocorrelated ($0\leq\gamma<1$) finite signals.
The edge values for $h(\pm 15)$ versus the autocorrelation exponent value $\gamma$ were investigated, keeping $L$ fixed. Examples of this dependence for $L=2^{12}$ and $L=2^{20}$ are shown in Fig.6. We found that cases for other lengths (not shown) look similarly and indicate the excellent linear decreasing function of $h(\pm 15)$ versus $\gamma$ in the whole range of autocorrelation exponent. Thus one gets:
\begin{equation}
\Delta h(\gamma,L) = A(L)\gamma + B(L)
\end{equation}
where the coefficients $A(L)$ and $B(L)$ depend on $L$ only. They can be further specified if the form of $\Delta h_1(L)$ and $\Delta h_0(L)\equiv \Delta h(\gamma=0, L)$ functions are used as boundary conditions.

The first boundary condition, i.e. $\Delta h_1(L)$, was already specified in Eq.(6). The profile of the second one ($\Delta h_0(L)$) can be deduced from  Figs. 6,7. The extrapolation of the fitting lines $h(\pm 15)$ versus $\gamma$ to the point $\gamma\rightarrow 0$  (see Fig.6) gives the collection of $\Delta h(0,L)$ values, plotted against the length of time series in Fig.7. This can be done for the central values as well as for the data satisfying $95\%$ confidence level. It is seen from the Fig.7 that for fully autocorrelated time series ($\gamma\rightarrow 0$) $\Delta h_0(L)$ is represented again by the power law:
\begin{equation}
\Delta h(0,L)=C_0L^{-\eta_0}
\end{equation}
with some constants $C_0$ and $\eta_0$ to be determined from the fit.

Linking the shape of boundary conditions given by Eqs.(6) and (9) with the general linear dependence in Eq.(8), one arrives with the final formula for $\Delta h(\gamma, L)$:
\begin{equation}
\Delta h(\gamma,L)=C_1 L^{-\eta_1}\gamma + C_0 L^{-\eta_0} (1-\gamma)
\end{equation}

The shape of the $95\%$ confidence level for multifractal background noise will be given by the same formula but with different coefficients calculated according to formulas like in Eq.(7). The final values of these coefficients are collected in Table 1.

\vspace{2em}
\begin{table}[!h]
\centering
\begin{tabular}{||c|c|c|c||c|c|c|c||} \hline
$C_1$ & $\eta_1$ & $C_0$ & $\eta_0$ & $C^{95\%}_1$ & $\eta^{95\%}_1$ & $C^{95\%}_0$ & $\eta^{95\%}_0$\\ \hline
0.603 & 0.175 & 0.453 & 0.124 & 0.631 & 0.171 & 0.484 & 0.120\\ \hline
\end{tabular}
\caption{The collected results for coefficients  of the fit in Eq.(10) describing the multifractal noise thresholds. Ensemble of $10^2$ independent realizations of time series was considered.}
\label{tab1}
\end{table}
Our results may also be presented graphically in a form of 'phase-like' diagrams (see Fig.8).Three separable areas in $(\Delta h, \gamma)$ plane can be distinguished for every $L$. The first area corresponds to multifractality connected entirely with finite size effects. It is marked in red in Fig.8. The second domain, marked in light green, is related to $(\Delta h, \gamma)$ range where multifractality may occur due to the long memory present in data but independent on the chosen time scale. The 'true' multifractality, i.e. the one related with long memory entirely dependent on the time scale, may occur only in the white region (at $95\%$ confidence level).

\section{Multifractal spectrum analysis of finite size effects.}

The multifractal spectrum width is considered as another useful measure of multifractality included in analyzed signal. Analogically to the \(\Delta h(L,\gamma)\) analysis, one may ask for the dependence of multifractal spectrum width \(\Delta\alpha\) on the signal length $L$ and on its persistency level $\gamma$. These results are obtained with use of Eq.(2) applied to previously discussed generalized Hurst exponent calculations. Thus, the results should lead to similar qualitative conclusions, nevertheless quantitatively they might be also valuable from practical point of view.

The examples of multifractal spectrum ($\alpha, f(\alpha)$) for finite monofractal signals are shown in Fig.9. Three cases: for strongly autocorrelated, medium autocorrelated, and weakly autocorrelated signals are considered there for two distinct lengths of data: $L=2^{12}$ and $L=2^{20}$.

As previously, the first step is to examine the $\Delta\alpha$ characteristics obtained for randomly shuffled signal ($\gamma=1$). Fig.10 shows the minimal $\alpha_{min}$ and maximal $\alpha_{max}$ value of $\alpha$ parameter revealing lack of dependence on $\gamma$. This proves again that shuffling was sufficient. The dependence of $\Delta\alpha(\gamma=1) \equiv \Delta\alpha_1$  on $L$ obeys a power-law relation:

\begin{equation}
\Delta\alpha_1(L)=D_1L^{-\xi_1},
\end{equation}
shown in Fig.11, with unknown constants $D_1$ and $\xi_1$. The 95\% confidence level is given by equation analogical to Eq.(7).

The edge values for $\alpha_{min/max}$ as a function of $\gamma$ are presented in Fig.12 for particular lengths $L=2^{12}$ and $L=2^{20}$, to confront them with previous findings for $h(\pm 15)$. The linear dependence for all other lengths (not presented) is also observed. Once we repeat the same approach as in the previous section to $\Delta\alpha(L,\gamma)$ dependence taking into account the second boundary condition $\Delta\alpha_0(L) \equiv \Delta\alpha(L,\gamma=0)$, we arrive with the final formula describing the character of multifractal spectrum width (see Fig.13), similar to the one found in Eq.(10):

\begin{equation}
\Delta \alpha(\gamma,L)=D_1 L^{-\xi_1}\gamma + D_0 L^{-\xi_0} (1-\gamma).
\end{equation}
The values of fitted parameters are gathered in Table 2.

\vspace{2em}
\begin{table}[!h]
\centering
\begin{tabular}{||c|c|c|c||c|c|c|c||}
\hline
$D_1$ & $\xi_1$ & $D_0$ & $\xi_0$ & $D^{95\%}_1$ & $\xi^{95\%}_1$ & $D^{95\%}_0$ & $\xi^{95\%}_0$\\ \hline
0.686 & 0.129 & 0.572 & 0.089 & 0.784 & 0.120 & 0.670 & 0.079\\ \hline
\end{tabular}
\caption{The results of the fit for coefficients in Eq.(12) done on ensemble of $10^2$ independent realizations.}
\label{tab2}
\end{table}
\vspace{2em}

\section{Concluding remarks}

We have shown qualitatively and quantitatively how multifractality arises from the finite size effects and (or) from autocorrelations not changing with the time scale being formed by the
specific autocorrelation exponent $\gamma$. This kind of multifractality, called by us 'multifractal noise', should be clearly distinguished from the 'real multifractality' caused by memory effects dependent on the time scale and thus leading to different scaling properties at various scales. We provided analytical formulas describing the multifractal noise threshold which turns out to be the power law function of time series length $L$ and the linear function of autocorrelation exponent $\gamma$. Our approach differs from the one presented in  Ref.\cite{zhou} where long memory in signals was produced from the induced power spectrum profile instead of the direct autocorrelation input between data resulting immediately from Eq.(3).

Two description methods for multifractality  were considered, i.e. the generalized Hurst exponents and the multifractal spectrum analysis. In both cases multifractal residual effect in monofractal finite signals has been found. We have shown that this effect measured by the spread of generalized Hurst exponent  or the width of multifractal spectrum grows linearly with autocorrelations level in time series and decays according to power-law with their length. We have estimated numerically the level of such multifractal noise and we captured it in simple analytical equations.

Finally, one should compare the obtained multifractal noise threshold with examples of the real multifractal data. We took them from finance because of common agreement that multifractality is a characteristic feature of financial markets. This problem is considered in Fig.14, where the simulated 'phase-like' diagram for data length $L\sim 2\times 10^3$ is shown, together with multifractal properties of various markets -- both for price indices and for volatilities. The particular length $L$ has been chosen as the average length of available data for analyzed markets. The multifractal properties for price indices were taken from \cite{zunino}, once the respective features for volatilities were originally calculated by us for the purpose of this paper and are based on historical data available in web \cite{yahoo}.

The multifractality for the volatility series is noticed. However, the presence of multifractality for the price index data is already not so obvious for all markets. One may find indices where multifractality comes indeed as a result of scaling properties changing with the time scale (e.g. Venezuela, Indonesia, China). Simultaneously, there are markets where the observed multifractality is generated mainly (e.g. Philippines, Taiwan, Germany) or even entirely (Ireland) by the finite size effects. In the case of Philippines, Taiwan, Thailand, Germany, Spain and Greece  almost $80\%$ of multifractality in price indices is related to such an effect.

 Thus, the multifractal properties of real financial data, may substantially (even in $80\%-100\%$) originate from the multifractal noise, what makes difficult in some cases to separate what main phenomenon is really responsible for the effects one observes. This confirms that multifractality is very tiny and delicate effect and one should be especially careful drawing far-reaching conclusions from the multifractal analysis in finance and in other areas. Our formulas are general enough to be applied also to other kind of real data in order to distinguish if and how their multifractal properties result from the pure multifractal origin.


\begin{figure}[p]
\includegraphics[width=15truecm]{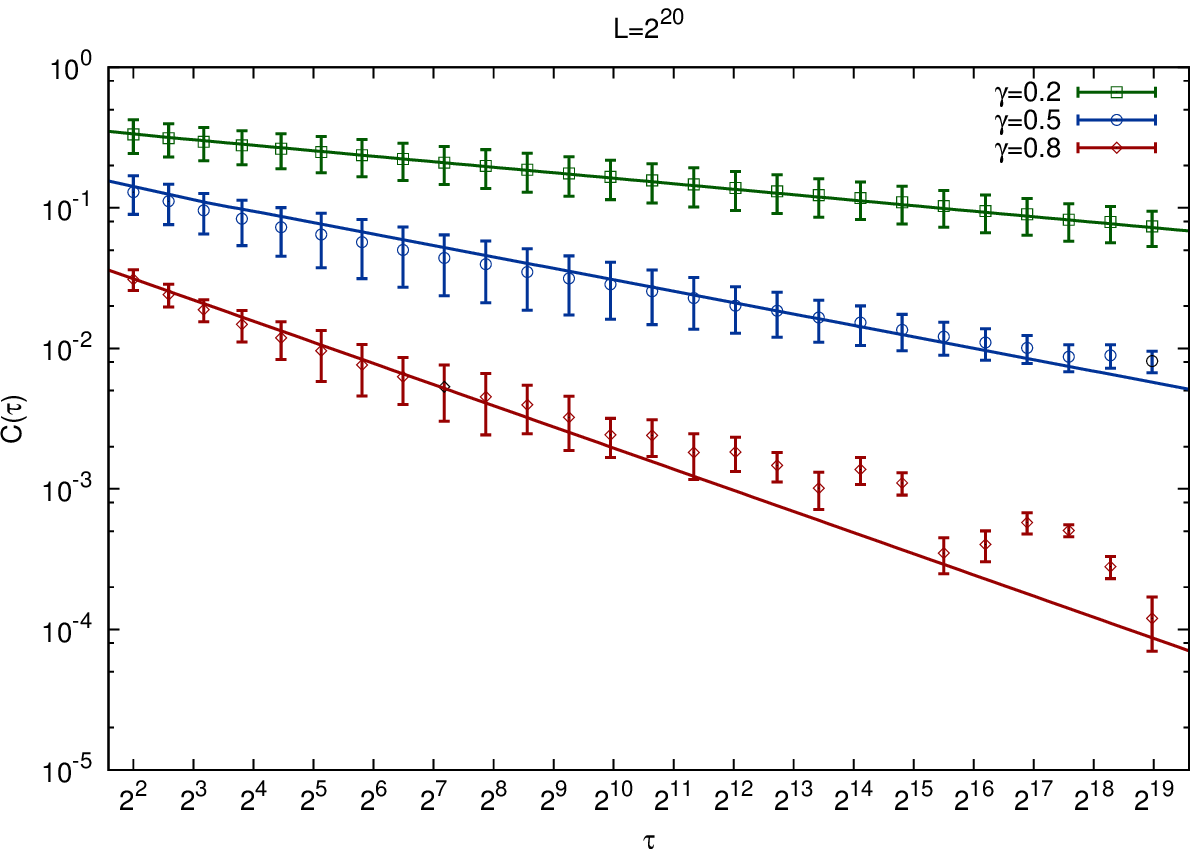}
\caption{Efficiency of FFM for replication of autocorrelation properties in time series. The examples for input values $\gamma=0.2$, $0.5$, $0.8$ are shown in log-log scale for the generated data of length  $L=2^{20}$. The lines present the fit to the desired power law dependence of Eq.(3), while error-bars show $1\sigma$ standard deviation following from the considered statistics of $10^2$ independent realizations. The output $\gamma$ values from the fit are found $\gamma_{out}=0.203 (\pm 0.009)$, $0.498 (\pm 0.012)$, $0.782 (\pm 0.054)$ respectively.}
\label{c-l}
\end{figure}

\begin{figure}[p]
\includegraphics[width=15truecm]{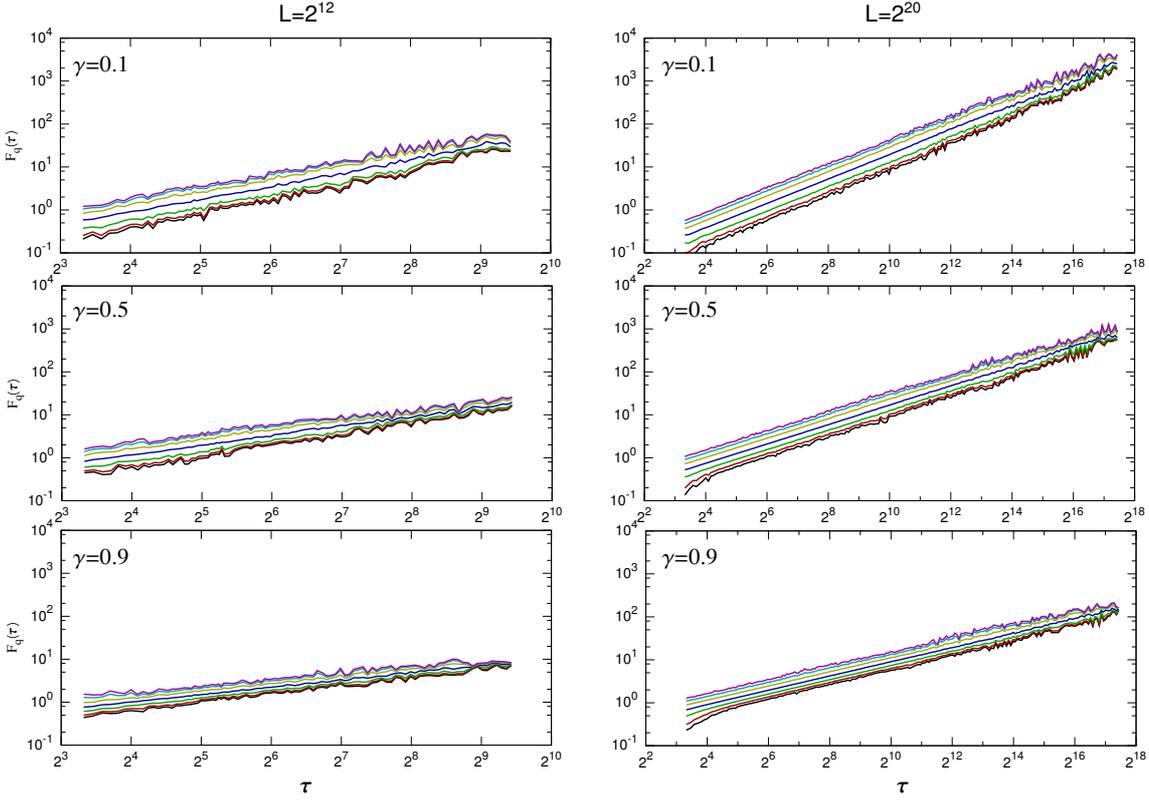}
\caption{Scaling of $q$-deformed fluctuations within MF-DFA. Results are presented for two different lengths of time series $L=2^{12}$, $2^{20}$, three autocorrelation parameters $\gamma=0.1$, $0.5$, $0.9$ and $q=-15,-10,-5,0,+5,+10,+15$. All plots confirm the proposed scaling range from $\tau=10$ till $\tau=L/4$.}
\label{Fq-tau}
\end{figure}

\begin{figure}[p]
\includegraphics[width=15truecm]{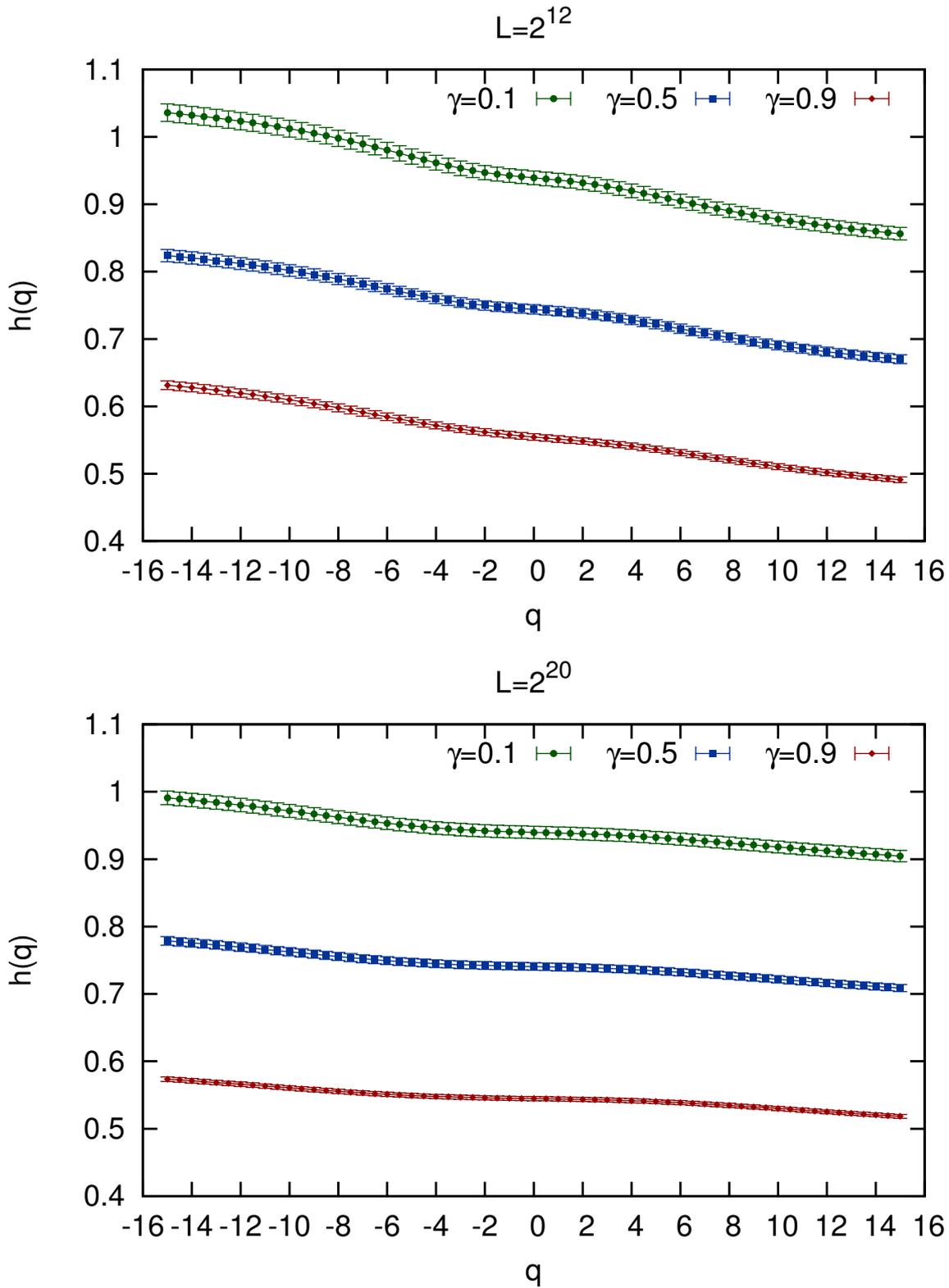}
\caption{Generalized Hurst exponent for monofractal signals generated within FFM with various autocorrelation properties. Two cases, for $L=2^{12}$ and $L=2^{20}$ are shown with different autocorrelation level. Error bars correspond to statistics of $10^2$ generated series.}
\label{h_sh-q}
\end{figure}

\begin{figure}[p]
\includegraphics[width=15truecm]{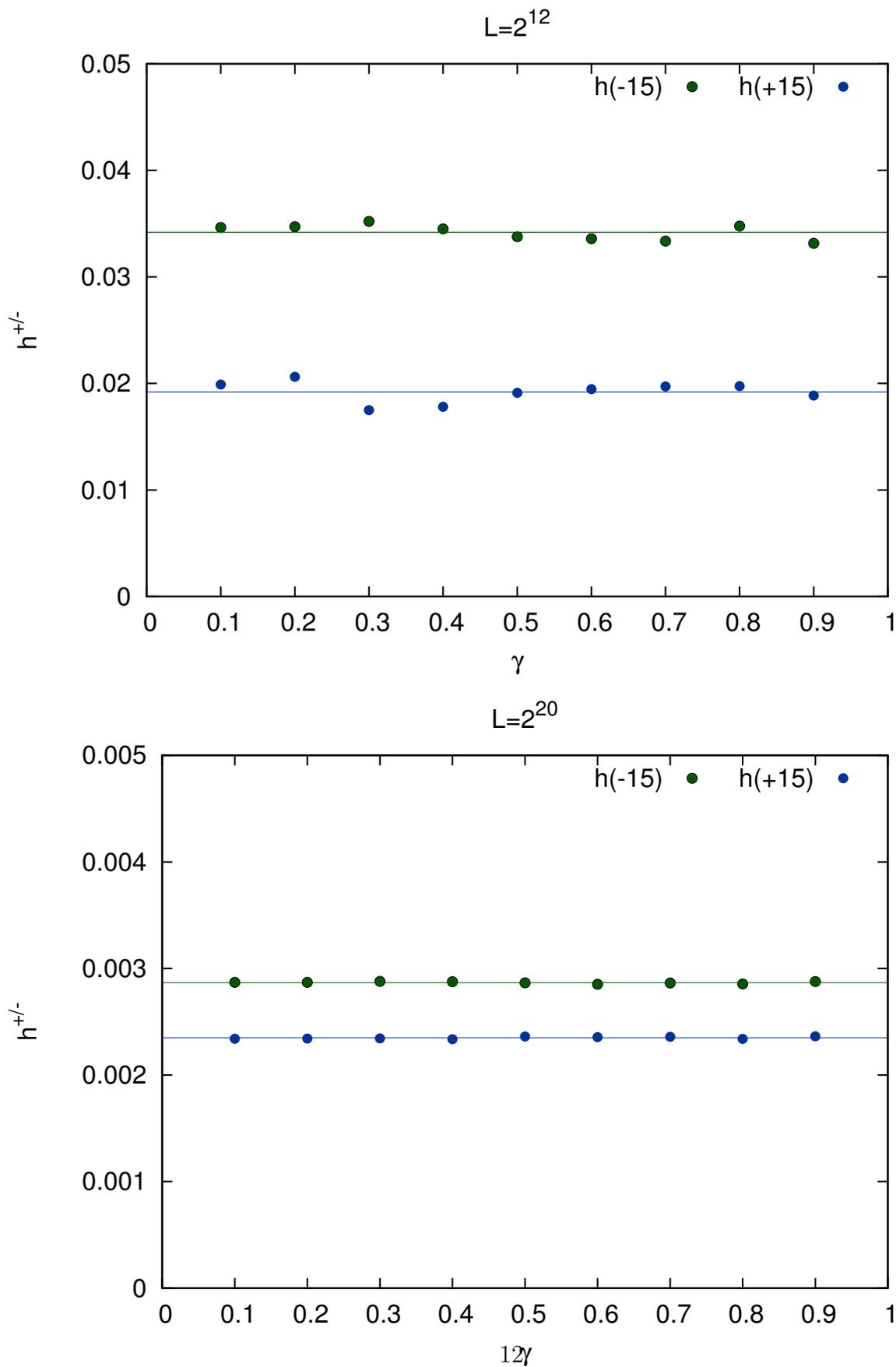}
\caption{Edge values of the generalized Hurst exponents $h(q)$ for two different lengths of time series $L=2^{12}$, $2^{20}$ constructed with long memory present ($\gamma<1$) and then shuffled to kill this memory. Dependence on the data length is readable.}
\label{h_edge_sh-gamma}
\end{figure}

\begin{figure}[p]
\includegraphics[width=15truecm]{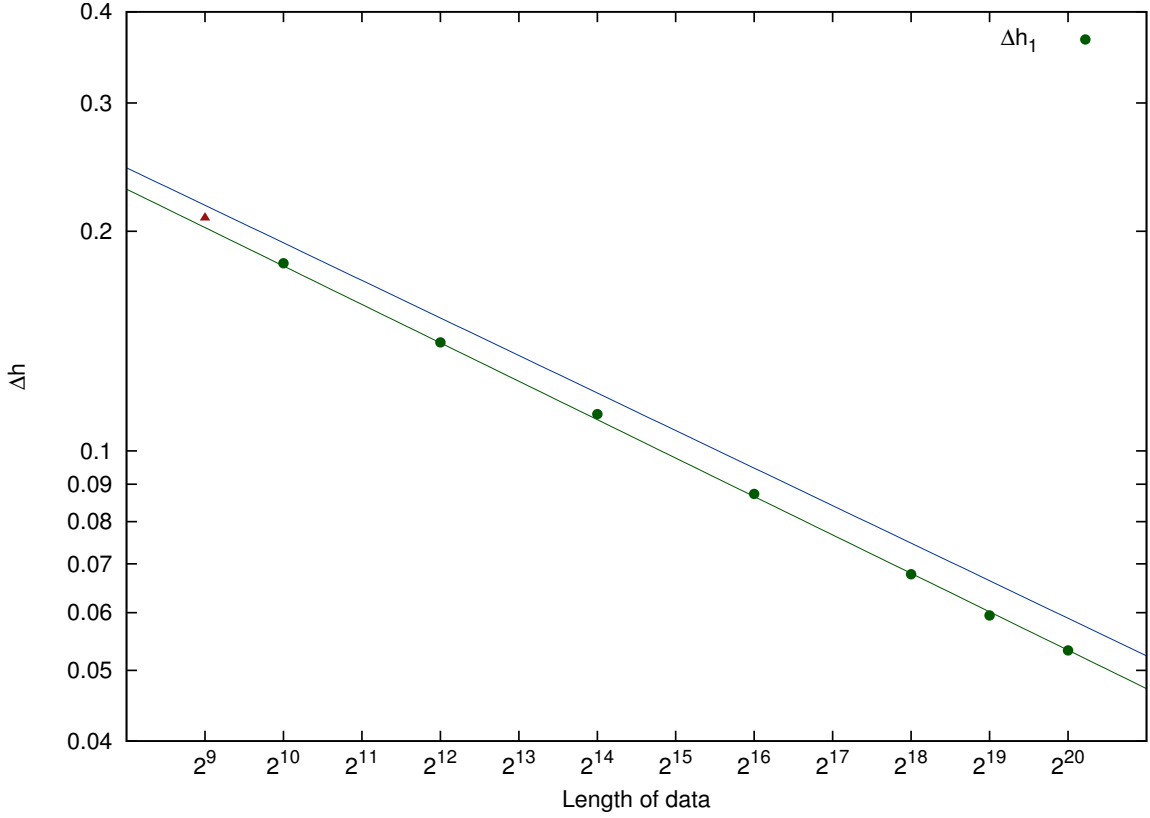}
\caption{Spread $\Delta h$ of generalized Hurst exponent  versus length of data drawn in logarithmic scale for the signal with no memory. Power-law dependence between  $\Delta h$ and the data length is visible. Data point corresponding to $L=2^9$ (marked as triangle) is slightly above the fitting line due to insufficient statistics for so short signal. Therefore, this point has been neglected during fitting procedure. Results of the fit are $C_1=0.603$, $\eta_1=0.175$ for the central values and $C_1^{95\%}=0.631$, $\eta_1^{95\%}=0.171$ for 95\% confidence level. The latter fit is marked as the blue top line.}
\label{h_sh-q}
\end{figure}

\begin{figure}[p]
\includegraphics[width=15truecm]{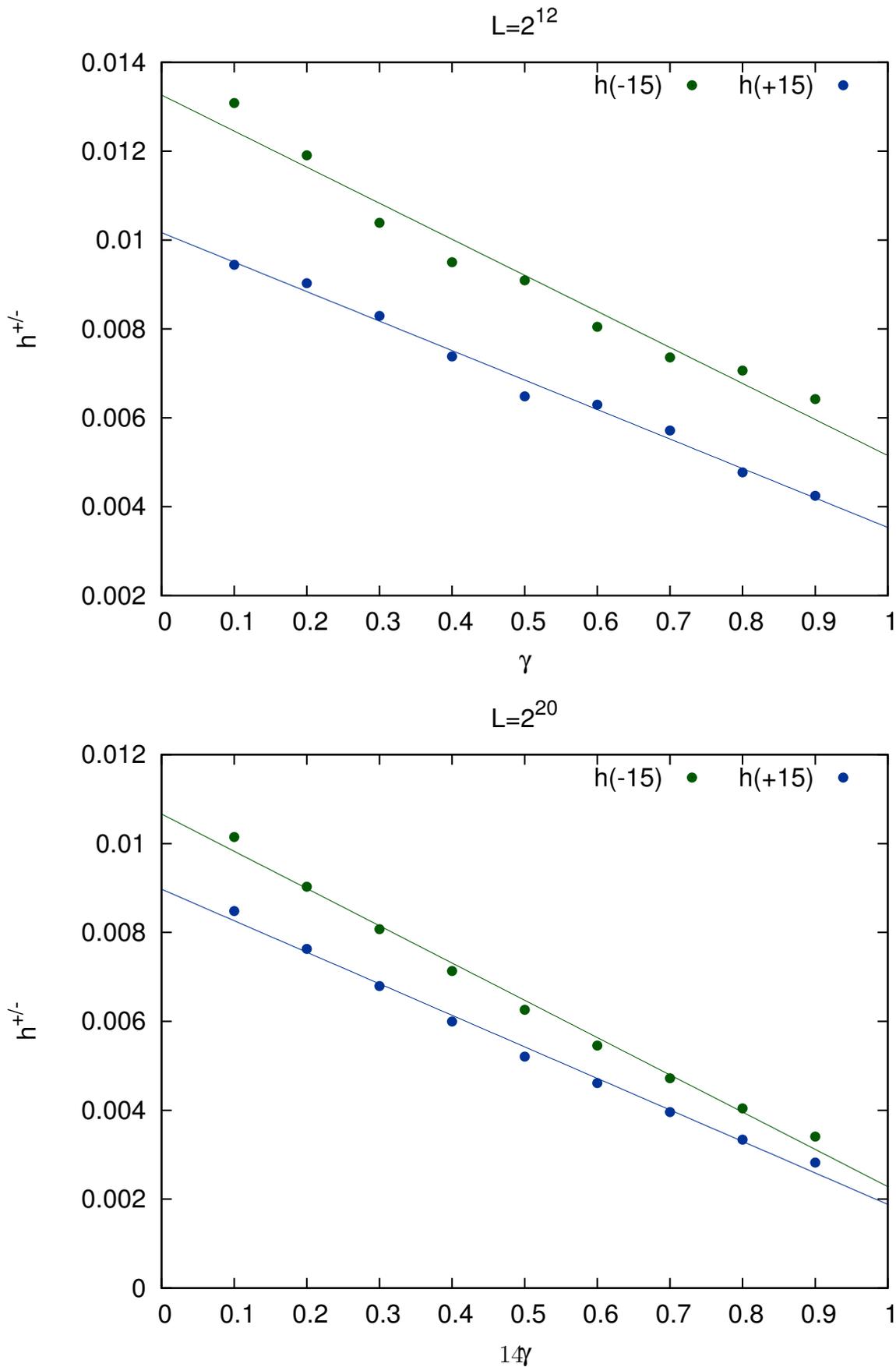}
\caption{Edge values of generalized Hurst exponent  for series with long-memory. Figures show the linear dependence between the edge values $h^{\pm}\equiv h(\pm 15)$ and $\gamma$ exponent. Extrapolation of fitted lines to the point $\gamma=0$ is interpreted as the edge values for fully autocorrelated signal ($C(\tau)\rightarrow 1,\ \forall\tau$)}
\label{h_sh-q}
\end{figure}

\begin{figure}[p]
\includegraphics[width=15truecm]{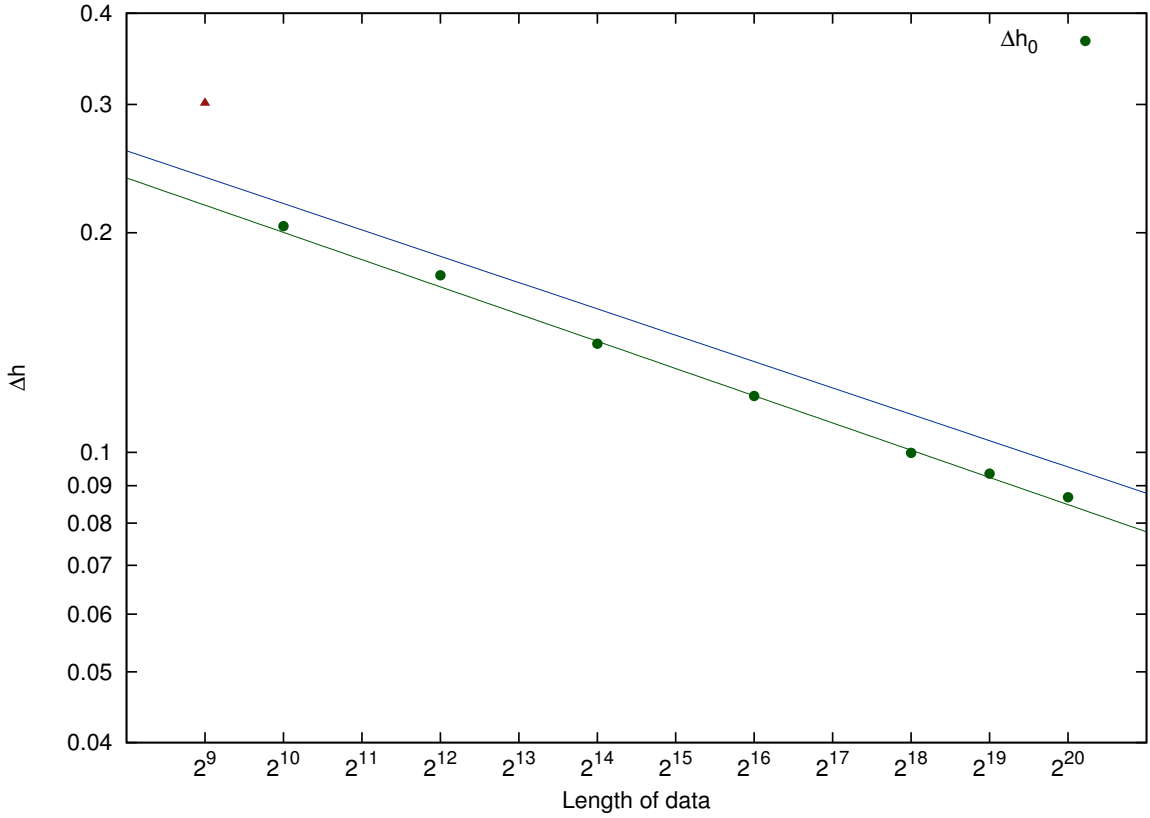}
\caption{Spread $\Delta h_0$ of generalized Hurst exponent  for fully autocorrelated time series ($\gamma=0$) versus the length of data. Green line presents the power-law  fit in log scale and the blue line corresponds to $95\%$ confidence level resulting from statistics. Fitted parameters are $C_0=0.453$, $\eta_0=0.124$ and $C_0^{95\%}=0.484$, $\eta_0^{95\%}=0.120$. Data point corresponding to $L=2^9$ has been removed from the fit due to insufficient statistics for so short signal leading to huge uncertainty in the estimation of generalized Hurst exponents within MF-DFA. }
\label{h_sh-q}
\end{figure}

\begin{figure}[p]
\includegraphics[width=15truecm]{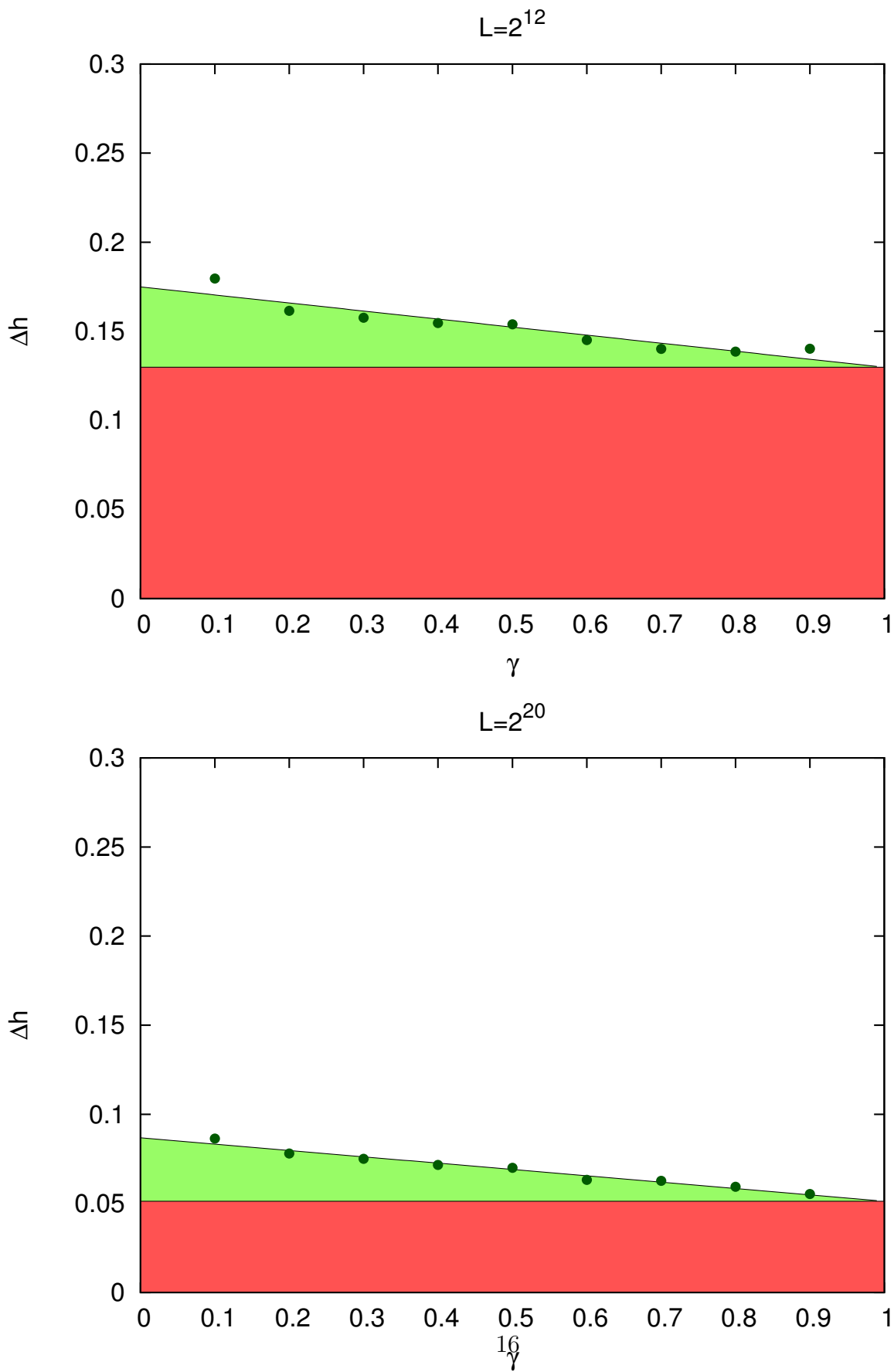}
\caption{Examples of phase-like diagrams for two data lengths $L=2^{12}$ and $L=2^{20}$. Three separable areas can be distinguished, marked in red, light green and white. They correspond to $(\Delta h, \gamma)$ domains where multifractality is caused at $95\%$ confidence level respectively by finite size effects, constant autocorrelation level independent on the time scale and 'true' multifractality (white).}
\label{h_sh-q}
\end{figure}

\begin{figure}[p]
\includegraphics[width=15truecm]{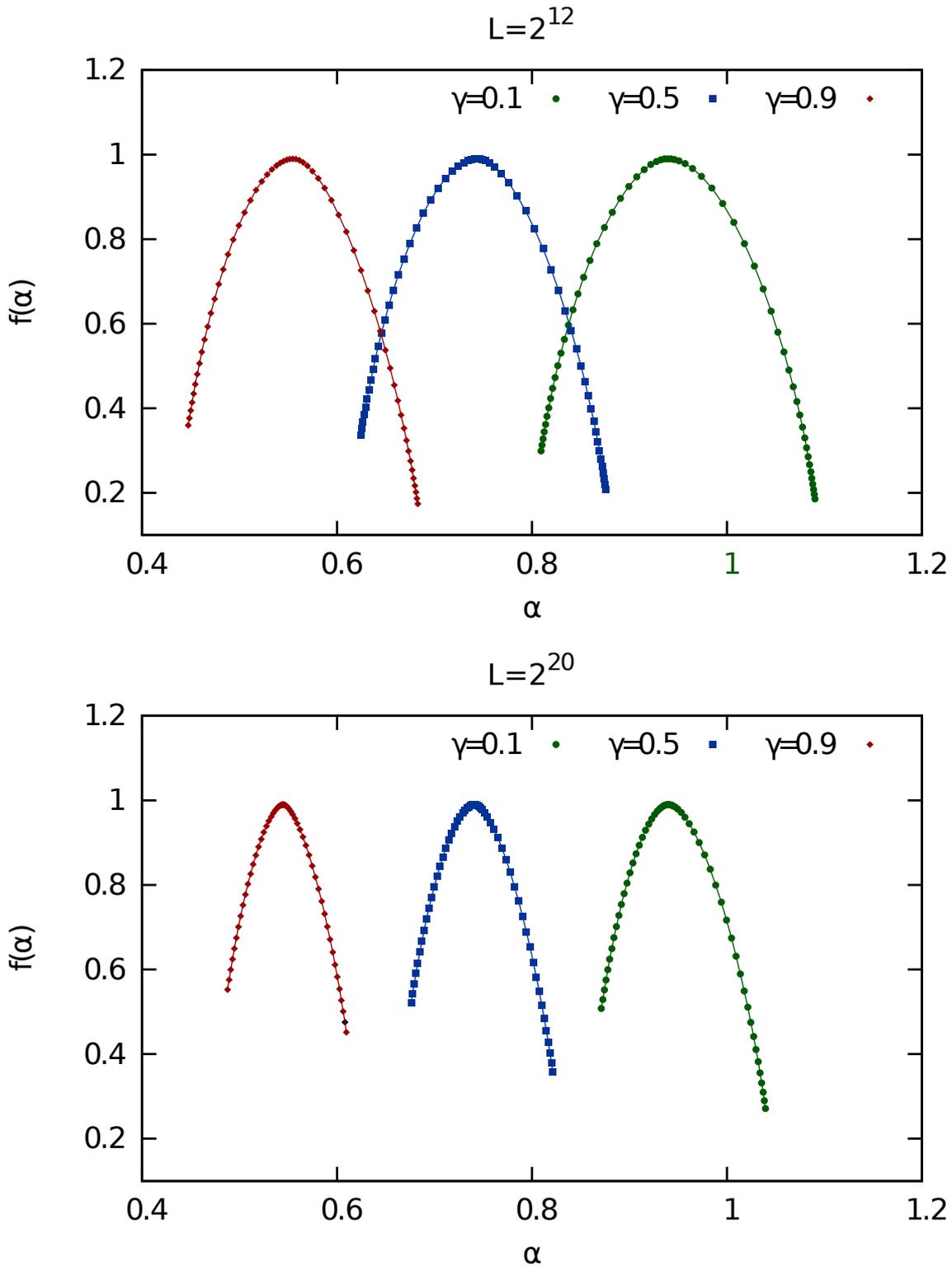}
\caption{Examples of multifractal spectra for finite monofractal strongly autocorrelated ($\gamma=0.1$), medium autocorrelated ($\gamma = 0.5$) and weakly autocorrelated ($\gamma=0.9$) signals. Mean results for two distinct lengths of signal $L=2^{12}$ and $L=2^{20}$ are compared for statistics of $10^2$ series.}
\label{f(alpha)}
\end{figure}

\begin{figure}[p]
\includegraphics[width=15truecm]{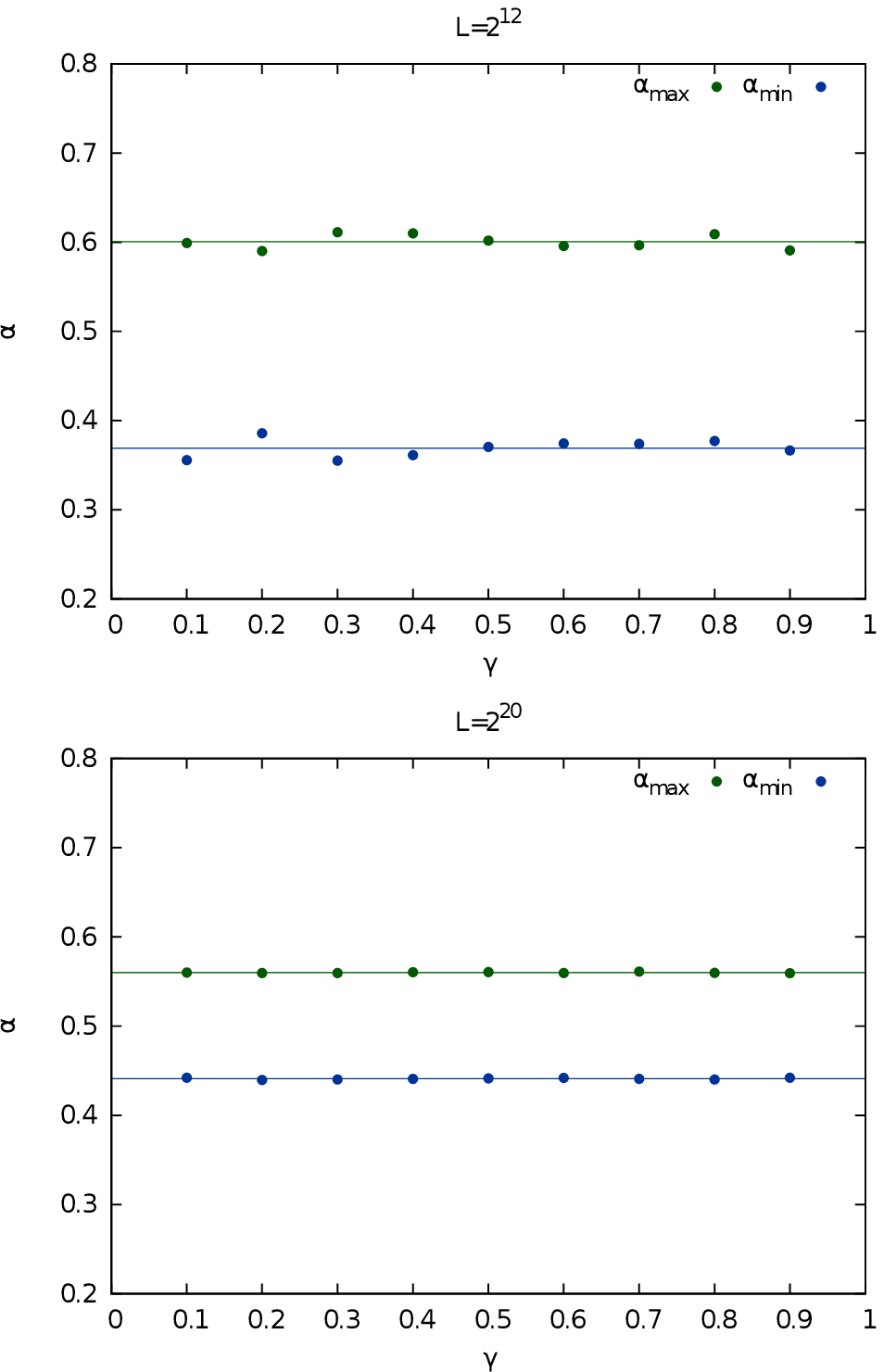}
\caption{Edge values of the H\"older parameter $\alpha$ for two different lengths of time series $L=2^{12}$, $2^{20}$ constructed with long memory present ($\gamma<1$) and then shuffled to kill this memory. Dependence on the data length is readable.}
\label{a_edge_sh-gamma}
\end{figure}

\begin{figure}[p]
\includegraphics[width=15truecm]{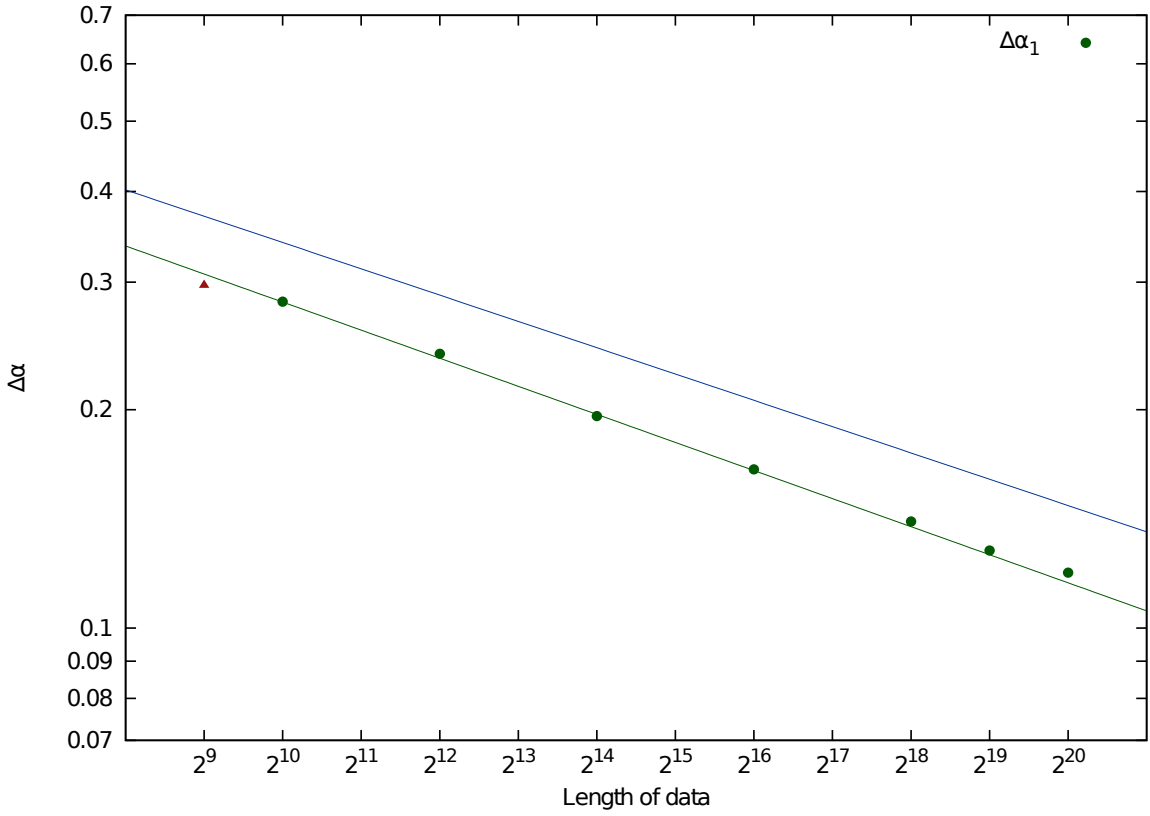}
\caption{Spread $\Delta\alpha_1$ of multifractal spectrum versus length of data drawn in logarithmic scale for the signal with no memory. Power-law dependence between  $\Delta\alpha_1$ and the data length is confirmed. Results of the fit are $D_1=0.686$, $\xi_1=0.129$ for the central values and $D_1^{95\%}=0.784$, $\xi_1^{95\%}=0.120$ for 95\% confidence level. The same notation applies as in Fig.5.}
\label{h_sh-q}
\end{figure}

\begin{figure}[p]
\includegraphics[width=15truecm]{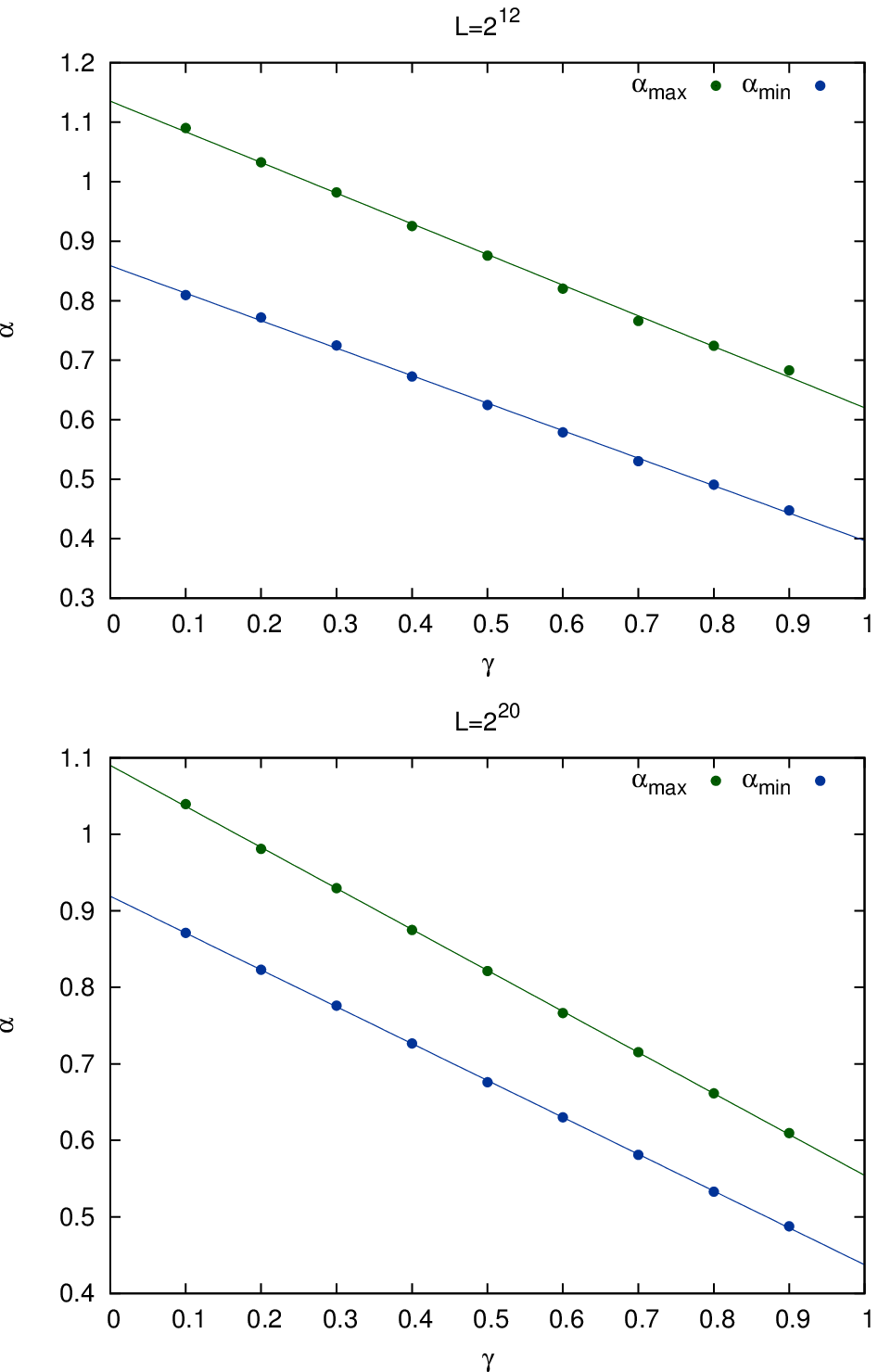}
\caption{Edge values of H\"older parameter $\alpha$ for series with long-memory. Figures clearly show the linear dependence between the edge values $\alpha_{min/max}$ and $\gamma$ exponent.}
\label{h_sh-q}
\end{figure}

\begin{figure}[p]
\includegraphics[width=15truecm]{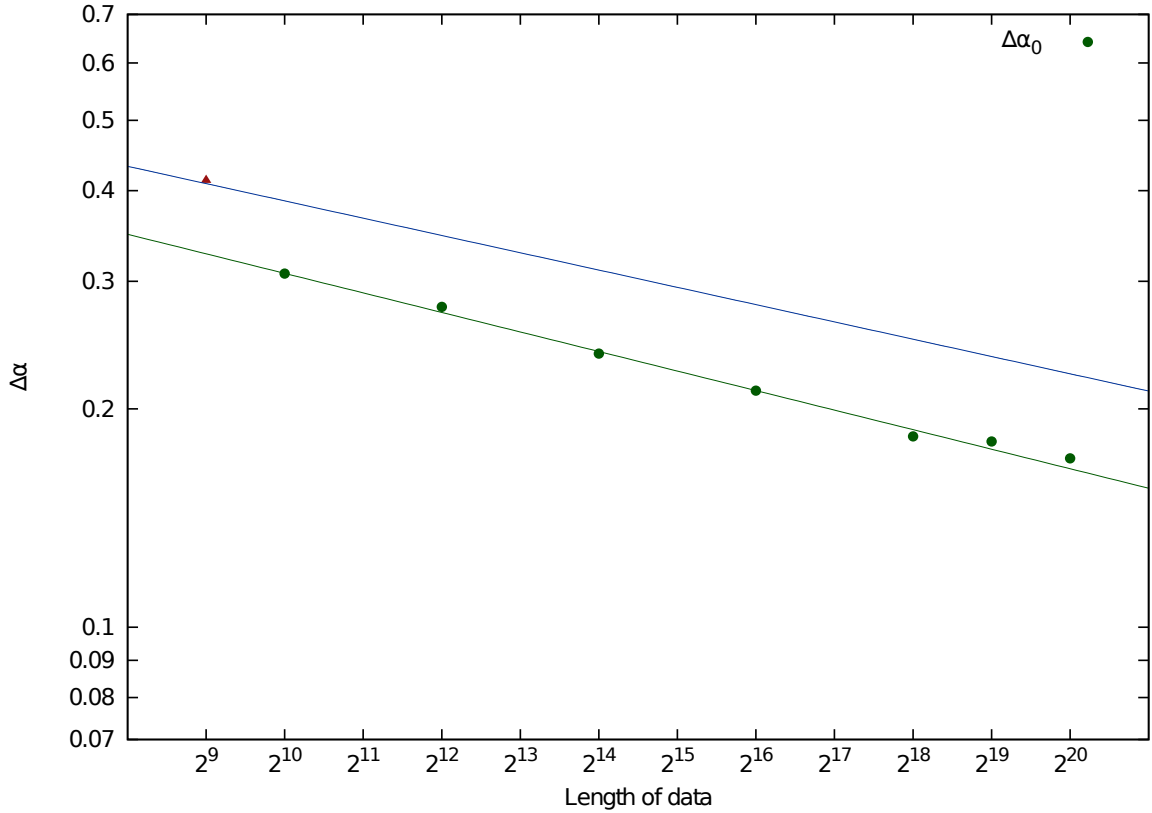}
\caption{Spread $\Delta\alpha_0$ of multifractal spectrum for fully autocorrelated time series ($\gamma=0$) versus the length of data. Fitted parameters are $D_0=0.572$, $\xi_0=0.089$ and $D_0^{95\%}=0.670$, $\xi_0^{95\%}=0.079$. Notation is inherited from Fig.7.}
\label{h_sh-q}
\end{figure}

\begin{figure}[p]
\includegraphics[width=15truecm]{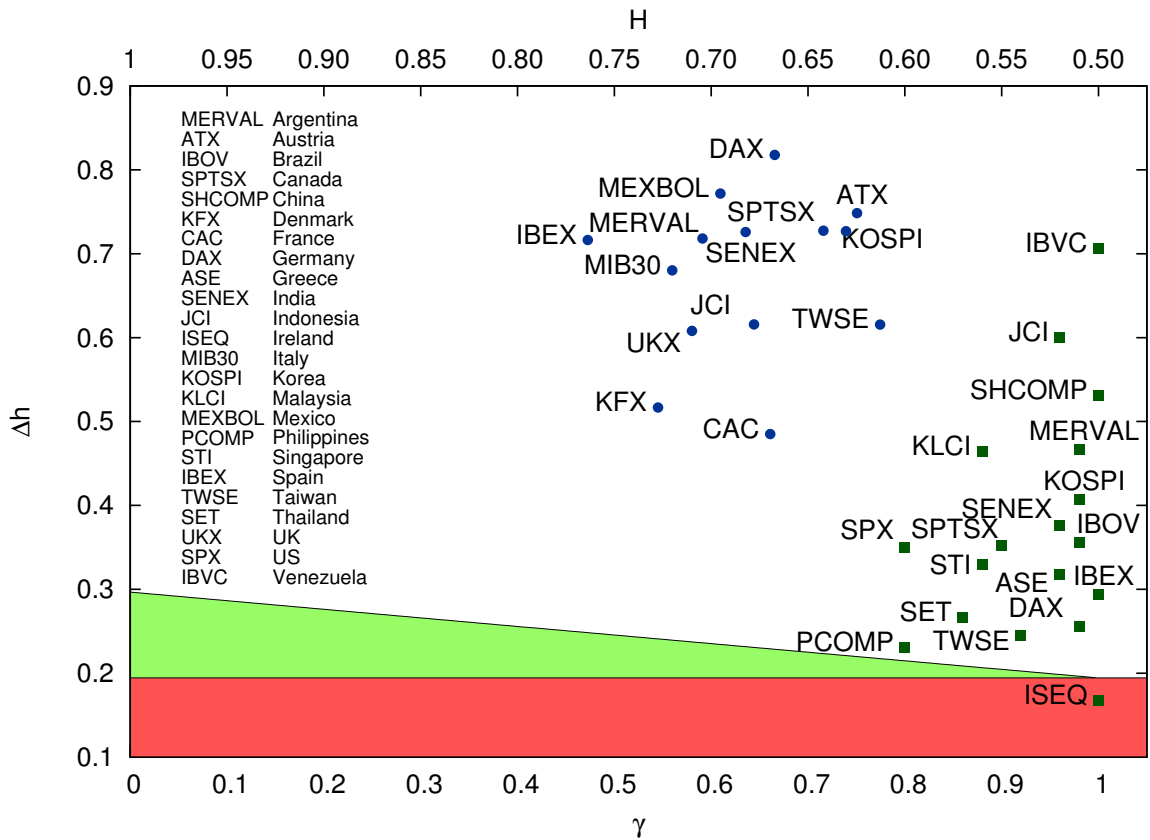}
\caption{Simulated 'phase-like' diagram for $L=2\times10^3$ with market data charted. Green squares represent multifractal properties of price indices and blue dots represent multifractal properties for volatilities. The Bloomberg code has been used to describe markets. The source of historical data is indicated in text. The corresponding values for Hurst exponents $H$ are also indicated (top axis) for convenience.}
\label{h_sh-q}
\end{figure}
\end{document}